# The Origins of Rydberg Atom Electrometer Transient Response and its Impact on Radio Frequency Pulse Sensing

Stephanie M. Bohaichuk, Donald Booth, Kent Nickerson, Harry Tai, James P. Shaffer

Quantum Valley Ideas Laboratories, 485 Wes Graham Way, Waterloo, ON N2L 6R2, Canada

*Rydberg atoms have shown significant promise as the basis for highly sensitive detectors of continuous radio-frequency (RF) E-fields. Here, we study their time-dependent response to pulse-modulated RF E-fields at 19.4 GHz using a cesium vapour cell at room temperature. We use density matrix simulations to explain the time scales that shape the transient atomic response under different laser conditions, finding them to be limited by dephasing mechanisms including transit time broadening, Rydberg-Rydberg collisions, and ionization. Using a matched filter, we demonstrate the detection of individual pulses with durations from 10 μs down to 50 ns and amplitudes from 15000 μV cm$^{-1}$ down to ~170 μV cm$^{-1}$, corresponding to a sensitivity of ~240 nV cm$^{-1}$ Hz$^{-1/2}$. Finally, we highlight the potential of a Rydberg vapour cell as a receiver by detecting pulse trains from a rotating emitter on a simulated passing aircraft.*

Rydberg atoms form a basis for several quantum technologies due to their uniformity, stability, and well-known properties [1]. For alkali atoms, like cesium (Cs), light can optically excite the outer electron into a Rydberg state that is sensitive to the presence of other atoms and external electric fields. There are many available excited states which offer significant tunability of Rydberg atom properties. One of the most promising applications is radio-frequency electrometry [2–6].

In these sensors, quantum interference between laser fields resonant with atomic transitions generates optical transmission in an absorbing vapour, termed electro-magnetically induced transparency (EIT) [7,8]. The presence of an RF electric field (E-field) resonant with another atomic transition disrupts the EIT and changes optical transmission in proportion to the RF E-field amplitude [2]. These sensors can be self-calibrated and detect electromagnetic fields across a broad MHz-THz frequency range, down to amplitudes of ~1 μV cm$^{-1}$ with a sensitivity of ~ μVcm$^{-1}$Hz$^{-1/2}$ [9–11], which can be improved to <55 nVcm$^{-1}$Hz$^{-1/2}$ by adding an auxiliary RF E-field [12]. Small dielectric vapour cells ($\lesssim$ 30 mm$^3$) minimally perturb the E-field and enable imaging with sub-wavelength spatial resolution [5,13,14].

While Rydberg sensors have been predominantly studied for sensing the amplitude of continuous wave RF E-fields, temporal dynamics can occur at sub-microsecond time scales [15,16]. This has made Rydberg sensors interesting for detecting modulated RF E-fields [16–20]. Furthermore, the polarization or phase of the RF E-field can be detected [12,21,22].

We study the atomic response to pulse modulated RF E-fields to demonstrate sensing of individual RF pulses of the type used in communications and radar, and to evaluate limitations on signal-to-noise ratio (SNR) and timing accuracy. Performance greatly improves by applying a matched filter tailored to the atomic response, despite the atomic system's nonlinearity and time variance. We detect RF E-fields down to ~170 μV cm$^{-1}$, corresponding to a sensitivity of ~240 nV cm$^{-1}$ Hz$^{-1/2}$ without the need for an auxiliary RF E-field or additional modulation. Finally, we demonstrate the use of a Rydberg sensor as a radar receiver for pulses emitted by a passing aircraft.

## Experimental Setup and RF Pulse Response

To study the response of Rydberg atoms to a pulsed RF E-field, we use a 3 cm long by 1cm square glass-blown cell filled with cesium vapour at room temperature. EIT in the vapour cell is generated using spatially overlapped counter-propagating IR probe (852.35 nm) and green coupling (509.31 nm) laser beams with ~140 μm and ~160 μm $1/e^2$ radii respectively. These are resonant with transitions between atomic states in the ladder system shown in Figure 1a, and are both polarized linearly, parallel to each other. In all measurements, the probe laser is offset locked to the Cs F=4 to F'=5 D2 transition using an external Fabry-Perot cavity and the Pound-Drever-Hall technique. The coupling laser is either scanned in frequency to measure steady-state EIT, or offset locked, using the same cavity, to resonance to measure a pulsed RF E-field. The laser spectral bandwidths are <100 kHz. Transmission of the probe laser through the vapour cell is detected using an avalanche photodetector with a bandwidth of 10 MHz. The detector output is fed into an oscilloscope, or an FPGA for further processing using a matched filter.

RF pulses are applied using pulse modulation of a synthesizer, with rise and fall times of < 80 ns, output to a RF horn antenna with a gain of ~15 dB placed ~25 cm from the cell. Unless otherwise specified, we use a frequency of 19.4 GHz, which couples the $55D_{5/2}$ and $53F_{7/2}$ Rydberg states of Cs, with a pulse repetition rate of 5 kHz. The RF E-field is linearly polarized, perpendicular to the laser polarizations and parallel to their propagation directions.

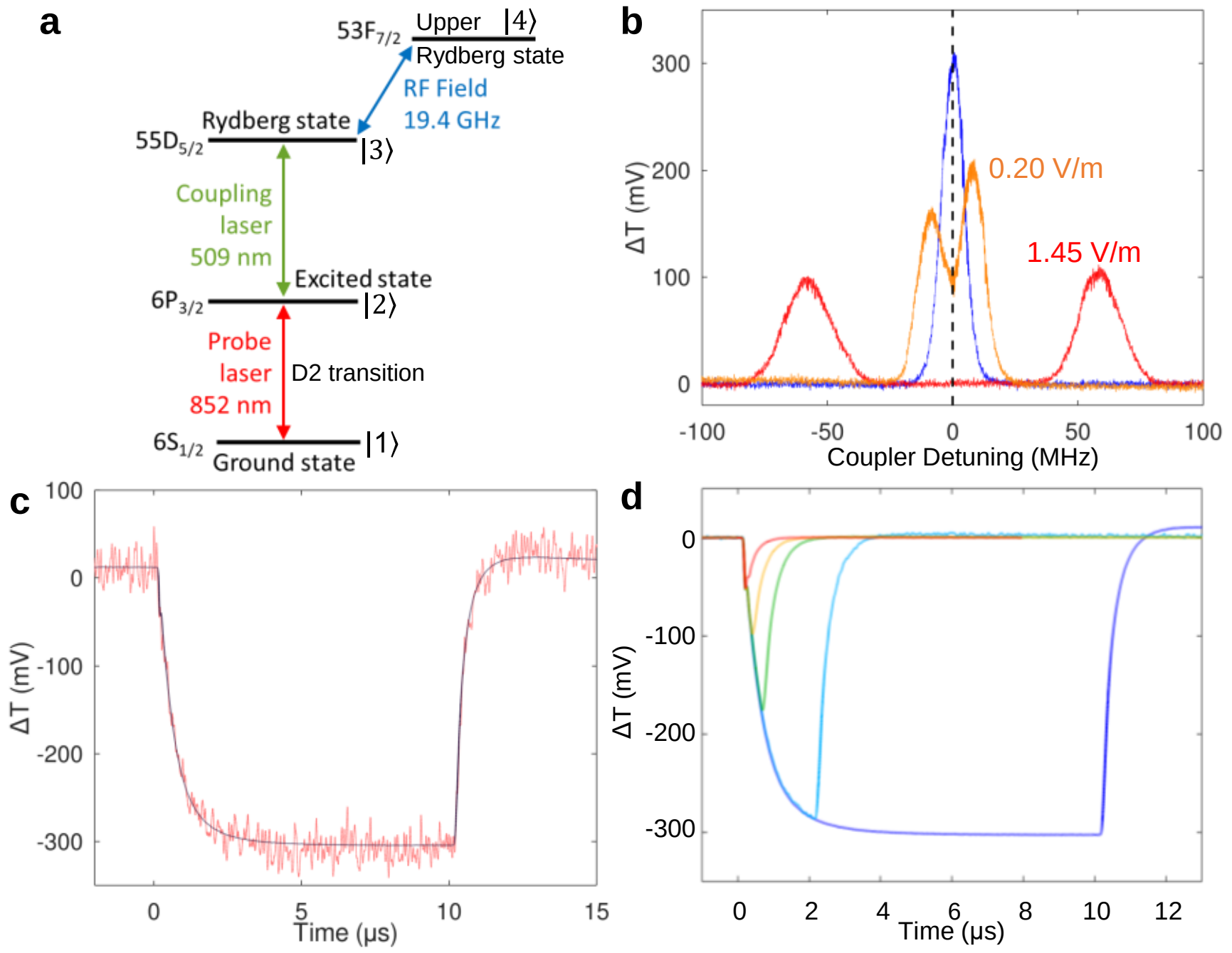

**FIG. 1** – (a) Energy level diagram for the experimental cesium system. (b) EIT peak obtained as the coupling laser is scanned across resonance. Autler-Townes splitting of the peak is shown in red and orange for two different continuous wave RF E-fields. To measure RF pulses, the coupling laser is locked on resonance (dashed line). (c) Transient atomic response to a 10 μs RF pulse, measured as a change in transmission of the probe laser, with both a single trace (red) and one averaged over $10^4$ cycles (purple) for an RF E-field of 1.45 V $m^{-1}$, corresponding to $\Omega_{RF} = 2\pi \times 117$ MHz. (d) Averaged atomic responses to RF pulses of varying pulse durations, ranging from 50 ns to 10 μs. Shorter pulses do not reach the full steady state depth, and as a result are harder to detect. $\Omega_P = 2\pi \times 3.5$ MHz and $\Omega_C = 2\pi \times 2.1$ MHz.

Figure 1b shows EIT measured from the vapour cell with no RF E-field applied (blue curve) as the coupling laser is scanned over the Cs $6P_{3/2} \leftrightarrow 55D_{5/2}$ transition. The full-width at half-maximum (FWHM) of the EIT peak at these laser conditions is ~11 MHz, and increases with the probe and coupling Rabi frequencies $\Omega_P$ and $\Omega_C$. Upon application of a continuous wave 19.4 GHz RF E-field, Autler-Townes splitting of the EIT peak is observed, shown in red and orange for two different field strengths. The slight asymmetry in peak height for the orange trace is an example of a weak background DC E-field present in the vapour cell. The peak splitting is given by $\nu = E\mu_{34}/h$ where $\mu_{34}$ = 6294.3$ea_0$ is the dipole moment of the $55D_{5/2} \leftrightarrow 53F_{7/2}$ transition, $h$ is Planck's constant, and $E$ is the amplitude of the RF E-field. If the RF E-field strength is sufficiently weak then splitting does not occur, but the amplitude of the on-resonance EIT signal decreases in what we refer to as the amplitude regime. In between these limits the peaks split but remain partially overlapped.

In the Autler-Townes splitting regime we can directly extract the RF E-field amplitude at the vapour cell using the known atomic dipole moment for the RF transition and the measured spectral splitting of the Autler-Townes peaks as the coupling laser is scanned over resonance. In the amplitude regime where the EIT peak splitting is not discernable in response to weak RF E-fields, <0.1 V $m^{-1}$, we extrapolate the RF E-field strength at the atoms in the vapour cell using a calibration factor. This calibration factor converts the nominal output E-field of the RF generator to an E-field received at the vapour cell, based on the linearity of the generator's attenuator and the dipole moment of the Rydberg-Rydberg transition. We determine this factor by measuring the splitting of the EIT peaks in response to continuous wave RF E-fields of various strengths within the Autler-Townes regime in comparison with the nominal RF generator output.

To detect an RF pulse, the coupling and probe lasers are locked on resonance (vertical dashed line in Figure 1b). Upon application of a RF pulse the EIT peak splits, resulting in a large drop in transmission, detected as a pulse shown in Figure 1c. The atomic response approximates the shape of the RF pulse envelope, but has slower leading and trailing edges which each take ~2 μs to reach steady state after an initial rapid <100 ns transient.

In Figure 1d, we examine the atomic response to various RF pulse durations down to 50 ns, the limit of the RF generator's capabilities. The pulses are comprised of the same atomic time scales, but the total depth is reduced as the RF pulse is shortened, resulting in a degraded SNR when detecting single pulses, as the atoms are unable to respond quickly enough to sub-2μs pulses to reach a steady state optical transmission. Despite this, it remains possible to detect very short pulses due to the rapid decrease in optical transmission on the leading edge.

## Modeling and Origins of Atomic Time Scales

To understand the origins of the atomic time scales we perform density matrix calculations, which follow the time-dependent master equation:

$$\dot{\rho} = \frac{2\pi i}{h}[\rho, H] + \mathcal{L}(\rho), \tag{1}$$

where $H$ is the Hamiltonian of the system, $\rho$ is the density matrix, $h$ is Planck's constant, and $\mathcal{L}$ is the Lindblad operator which describes decays and dephasing. The levels in the model system correspond to those shown in Figure 1a, with $|1\rangle$ referring to the ground state, $|2\rangle$ to the intermediate excited state, $|3\rangle$ to the lower Rydberg excited state, and $|4\rangle$ to the Rydberg excited state which is coupled to $|3\rangle$ by the time-dependent RF Rabi frequency. We add a fifth level, $|5\rangle$, as a dark state to represent atomic states that are not optically coupled to the primary system. This level better models the pulse time scales at higher Rydberg populations and optical Rabi frequencies, but is not necessary at low Rydberg populations. We describe the Hamiltonian of this system in the interaction picture by:

$$H = \frac{h}{2\pi}\begin{pmatrix} 0 & \frac{\Omega_p}{2} & 0 & 0 & 0 \\ \frac{\Omega_p}{2} & -\Delta_2 & \frac{\Omega_c}{2} & 0 & 0 \\ 0 & \frac{\Omega_c}{2} & -\Delta_3 & \frac{\Omega_{RF}(t)}{2} & 0 \\ 0 & 0 & \frac{\Omega_{RF}(t)}{2} & -\Delta_4 & 0 \\ 0 & 0 & 0 & 0 & 0 \end{pmatrix}. \quad (2)$$

$\Omega_P$, $\Omega_C$, and $\Omega_{RF}(t)$ are the Rabi frequencies of the probe laser, the coupling laser, and the RF E-field respectively. The atomic dipole moments used to determine the optical Rabi frequencies are averaged over the accessible hyperfine transitions in our setup, taking into account the laser polarizations. The RF E-field is initially off ($\Omega_{RF}$=0) for the first 20 μs so that the simulation reaches equilibrium, then the RF E-field is pulsed on for a 10 μs duration, with a 25 ns rise and fall time. The detuning of the $|2\rangle$ state is given by $\Delta_2 = -\Delta_P + k_P v$ and the detuning of the $|3\rangle$ state by $\Delta_3 = -\Delta_P - \Delta_C + (k_P - k_c)v$. Here, both lasers are locked on resonance so $\Delta_P = \Delta_C = 0$. $k_P$ and $k_C$ are the wavevectors of the probe and coupling lasers, respectively, while $v$ is the atomic velocity along the direction of the probe laser used to account for Doppler shifts. We obtain Doppler averaged values by integrating $v$ over the Boltzmann distribution:

$$\rho_{21} = \int \sqrt{\frac{m}{2\pi k_B T}} \exp\left(-\frac{mv^2}{2k_B T}\right) \rho_{21}(v)\, dv, \quad (3)$$

where $T$ is the vapour cell temperature, $m$ the atomic mass of the alkali atom used, here $^{133}$Cs, and $k_B$ is Boltzmann's constant. We use the imaginary part of the density matrix element $\rho_{21}$, which is proportional to the absorption coefficient, $\alpha$, to obtain the response of the vapour. This closely approximates the total change in intensity of the probe laser in the vapour cell under weak absorption conditions, as an absolute value for total absorption can be difficult to model due to uncertainties in the atom number density, optical losses including reflections and absorption due to the vapour cell walls and downstream optical components, and detector sensitivity and gain.

In the Lindblad operator, we include $\Gamma_{21} = 2\pi\times5.2$ MHz as the well-known decay rate from $6P_{3/2}$ to $6S_{1/2}$ [23], and $\Gamma_{32} = 2\pi\times10.4$ kHz as a radiative decay rate for the Rydberg state to the excited state. $\Gamma_{31} = \Gamma_{41} = \Gamma_{51} = 2\pi\times275$ kHz represents a transit time of atoms through our laser beams. When Rydberg population is low, a dark state is not needed to reproduce the pulse shape (i.e., $\Gamma_{35} = 0$). At higher Rydberg populations adding a finite dark state generation rate produces better fits. Overall, depending on the experimental conditions, we use $\Gamma_{35} = 2\pi\times$ 0 – 800 kHz. The best quantitative value depends on the optical Rabi frequencies used and is discussed in the Supplemental Material.

Simulated and experimental atomic responses to a 10 μs RF pulse at low optical Rabi frequencies ($\Omega_P = 2\pi\times2.0$ MHz, $\Omega_C = 2\pi\times0.6$ MHz) are shown in Figure 2a. The leading edge consists of two time-scales: a sharp decrease in transmission occurring over ~100 ns, followed by a slower exponential reduction in transmission over a few microseconds. During the initial transient, the two-level system on the D2 transition rapidly drives itself to equilibrium, based on the populations in the $6P_{3/2}$ and $6S_{1/2}$ states and their coherences, in response to effective alteration of the EIT by the RF E-field. The depth of the transient diminishes as $\Omega_P$ increases, shown Figure 2b. Moving further from the weak-probe limit to higher $\Omega_P$ results in more atoms initially populating the $6P_{3/2}$ state and less change due to the presence or absence of the coupling laser, so that when the RF E-field is turned on there is less change in $6P_{3/2}$ population. At larger $\Omega_P$, the pulse is dominated by the slower time constant. For fast timing applications, it is advantageous to use lower $\Omega_P$ where the sharp transient is deeper.

The slower dynamics are explained by the repopulation of the interaction region in the vapour cell due to atomic motion, set by the transit time of atoms through the beam and the populations of Rydberg and dark states. Prior to the RF pulse, EIT produces a steady state Rydberg population. A fraction of these end up in a dark state due to Rydberg-Rydberg collisions, radiative decay, and black-body radiation and ionization, and are unable to participate in the optical dynamics. When the Rydberg population is small, as in Figure 2a-b, inclusion of a dark state is not necessary to produce a good fit to experimental data. However, including it in the model better matches the pulse shapes when larger Rydberg populations are excited. Once the RF pulse is turned on, disrupting EIT, creation of Rydberg population is hindered as is any associated dark state population. The atoms present in the upper Rydberg state, which becomes populated when the RF is on, the lower Rydberg state, and the dark state slowly drift out of the laser beams and are replaced by ground state atoms, at a rate set by the transit time. An exponential fit to the portion of the pulse after the initial transient yields an effective time constant that increases linearly when we increase the laser beam diameter and therefore the transit time (see Supplemental Material). The use of large laser beams may not lead to an increased SNR when detecting short pulses, as it is not conducive to quickly reaching equilibrium depth given that the transit time dominates the overall atomic response time.

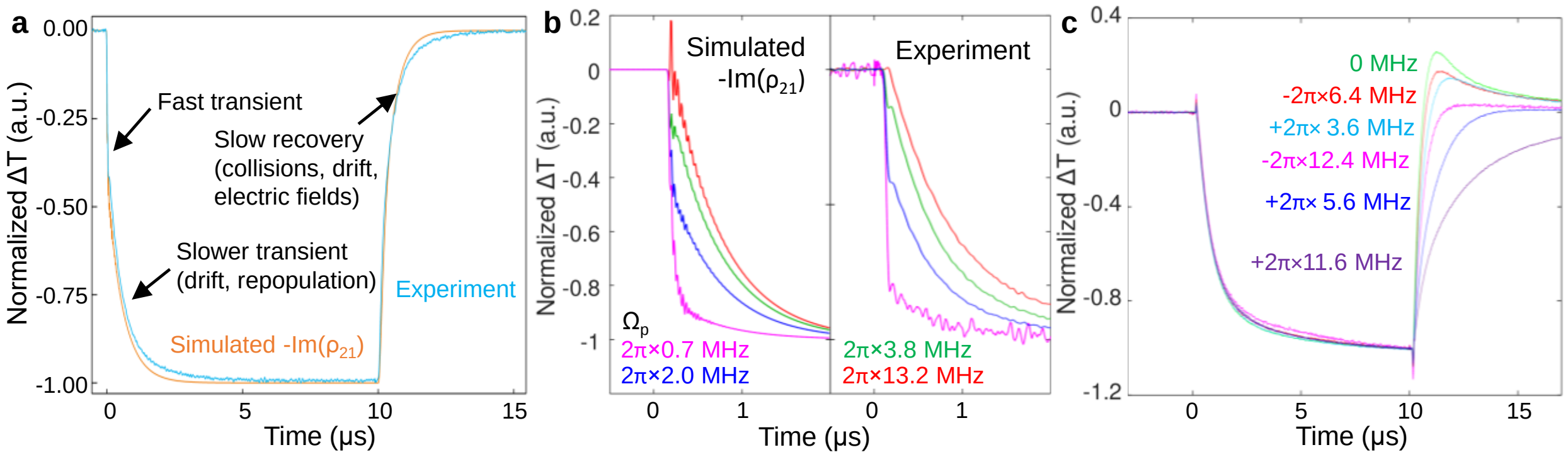

**FIG. 2** – (a) Experimental atomic response to a 10 μs RF pulse (blue) compared to one simulated with a density matrix model (orange). The pulses consist of several atomic time scales which are labelled by their origins. $\Omega_P = 2\pi\times2.0$ MHz, $\Omega_C = 2\pi\times0.6$ MHz, and $\Omega_{RF} = 2\pi\times60$ MHz. (b) Changes to the leading edge of the atomic response with the probe power, shown in both modeling (left panel) and experiment (right panel). $\Omega_C = 2\pi \times 0.6$ MHz, $\Omega_{RF} = 2\pi \times 60$ MHz. (c) Coupling laser detuning has a strong impact on the tail of the experimental response to an RF pulse. $\Omega_P = 2\pi \times 10.2$ MHz, $\Omega_C = 2\pi \times 2.2$ MHz, $\Omega_{RF} = 2\pi \times 119$ MHz. All experimental pulses are averaged $>10^3$ cycles.

Recovery of the pulse's trailing edge can be longer than the dynamics of the leading edge, and at higher laser powers often displays enhanced transmission. We attribute this in part to collisional-dependent ionization and electric field effects in the vapour cell which can take longer to re-equilibrate, up to ~100 μs (see Supplemental Material), after changes to the Rydberg population occur during and shortly after the pulse. These effects are especially apparent as the coupling laser is detuned, as shown in Figure 2c, impacting the overshoot and recovery time of the pulse's trailing edge. Detuning the coupling laser causes excitations of Rydberg atom pairs with different forces of attraction or repulsion, i.e., gradients of pair potential curves, resulting in different collisional and ionization rates depending on the amount and sign of the detuning. Furthermore, we observe that the long transient tail depends on the position of the lasers relative to the vapour cell walls, suggesting that background electric fields in the vapour cell can also be influential (see Supplemental Material).

## Detection of Individual RF Pulses

Figure 3a shows the measured atomic response to 19.4 GHz pulses of different amplitudes, with the corresponding experimental EIT peak splitting from continuous RF E-fields shown in the inset. The primary change is the depth of the pulse, as lower RF amplitude leads to reduced peak splitting.

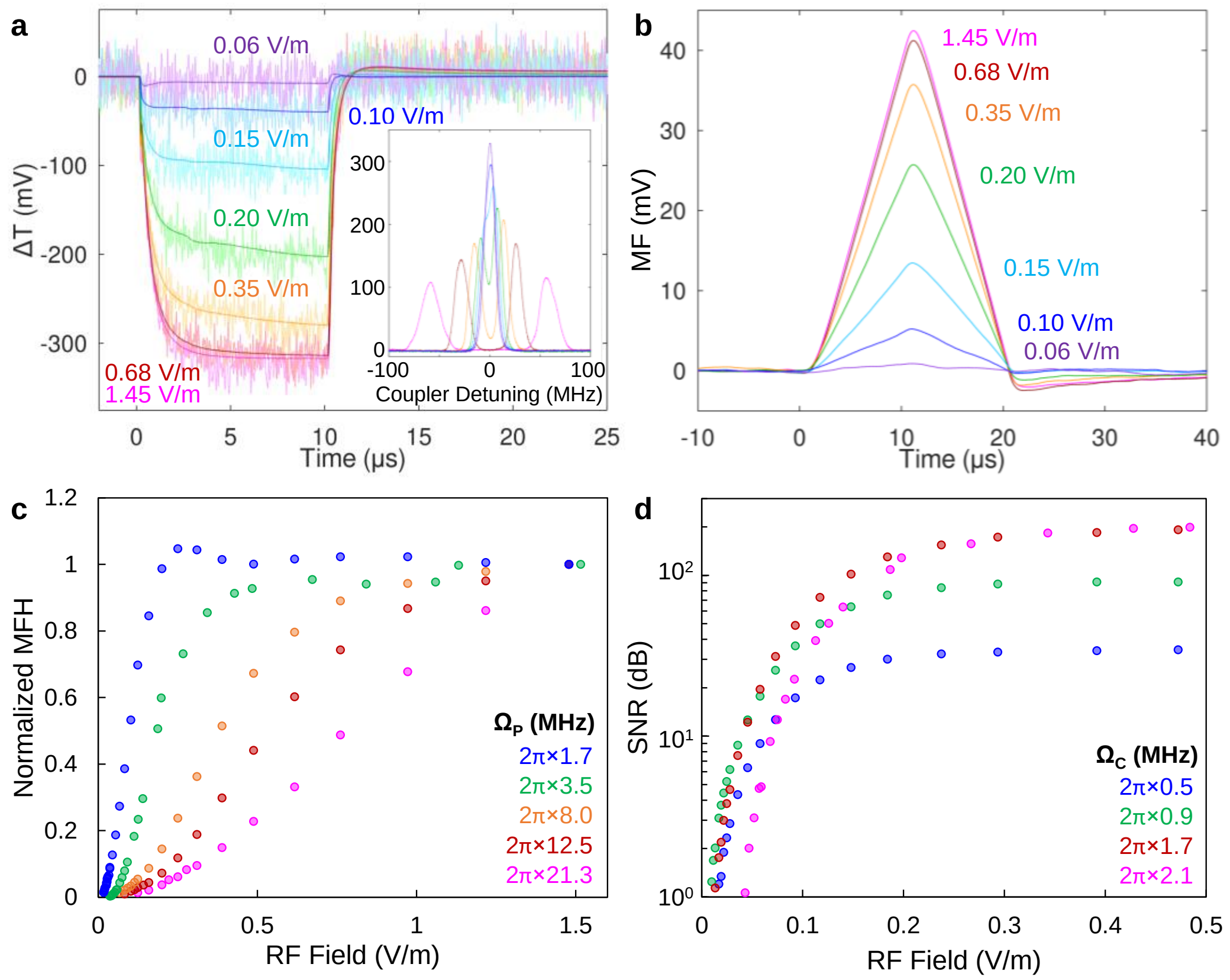

**FIG. 3 –** (a) Experimental atomic responses to a 10 μs RF pulse of various E-field amplitudes, shown as both single traces (faint lines) and traces averaged over $10^4$ cycles (darker lines). The inset shows the corresponding Autler-Townes splitting of the EIT peaks in response to a continuous wave RF E-field. $\Omega_P = 2\pi \times 3.5$ MHz, $\Omega_C = 2\pi \times 2.1$ MHz. (b) Output of the matched filter (MF) when applied to pulses shown in (a). The peak of the MF gives the pulse timing. (c) Peak height of the matched filter signal (MFH) as a function of probe laser power, normalized to the value at full EIT peak splitting (large RF E-field) at fixed $\Omega_C = 2\pi \times 2.4$ MHz. (d) Signal-to-noise (SNR) ratio of the matched filter as the coupling laser power is varied with fixed $\Omega_P = 2\pi \times 3.5$ MHz.

To improve the detection of weak RF pulses we apply a matched filter, which is used in many signal processing applications to extract a known pulse shape from white noise [24]. We implement the matched filter on an FPGA to perform real-time analysis. The output peak corresponds to the time of maximum cross-correlation, i.e. best match, of an expected pulse template to a measured noisy waveform. The template is obtained from experimental data taken at the same laser conditions, averaged over cycles to reduce noise. A simulated template can be used instead, though given that collisions and cell-dependent electric fields can be non-trivial to accurately capture, better performance of the matched filter is obtained at higher optical powers by using an experimental template. We cannot use the outgoing RF pulse as the expected template, in contrast to many implementations, as the atomic response shape does not match the square RF pulse envelope due to the finite atomic response time. As the pulse shape varies minimally with RF amplitude, we use the same pulse template at different RF amplitudes. We address the consequences of template-experiment mismatch later. Another difference between traditional receivers and a Rydberg atom receiver is that we have directly coupled the FPGA to the sensor without an intermediate filter or amplifier.

Figure 3b shows the output of the matched filter applied to individual pulses under the same conditions as in Figure 3a. The matched filter significantly suppresses noise, allowing the timing of weaker amplitude RF pulses to be extracted. The timing extracted by the matched filter is the arrival time of the pulse plus its duration as determined by the length of the template, here 11.1 μs.

As the EIT peak FWHM is strongly influenced by the combination of probe and coupling Rabi frequencies, we vary them to determine the optimal conditions for detecting weak RF E-fields. Because total absorption scales with probe laser intensity, larger $\Omega_P$ typically leads to larger pulse depths and therefore larger matched filter signals. However, our photodetector saturates at large probe powers, requiring the addition of a neutral density filter in front of the detector. Thus, for better comparison in Figure 3c we have normalized the matched filter peak height at a given $\Omega_P$ to its value at large RF E-fields, both averaged over 300 cycles. This maps changes in EIT peak width and overlap as a function of RF amplitude and $\Omega_P$, for fixed $\Omega_C = 2\pi \times 2.4$ MHz. At small $\Omega_P$, the EIT peaks are spectrally narrower, so that the Autler-Townes regime is extended and a lower RF E-field can be reached before the peaks begin to overlap and pulse amplitude decreases. If the aim is to maximize sensitivity to weak RF E-fields then lower $\Omega_P$ is desirable. In contrast, if the aim is to differentiate a wide range of RF E-field amplitudes, then larger $\Omega_P$ is preferable because the pulse depth varies gradually with RF E-field amplitude due to broad overlapping EIT peaks.

Figure 3d shows changes in SNR as $\Omega_C$ is varied while $\Omega_P = 2\pi \times 3.5$ MHz. For analysis of the matched filter's performance, we define the signal-to-noise ratio (SNR) by taking an average of matched filter peak heights from 300 pulse traces then dividing by the standard deviation of noise measured from 300 traces of identical length taken with no RF E-field present. If a pulse could not be identified within a given trace, *i.e.* there was a missed detection or larger false alarm from a mistimed noise spike, then we set the detected height for that trace to zero. Each trace was at least 10X longer duration than an individual pulse to capture false alarms and provide a more realistic SNR. Reducing $\Omega_C$ improves the SNR at the lowest RF E-fields by spectrally narrowing the EIT peaks, but comes at the cost of reduced SNR at high RF E-fields due to lower EIT peak amplitude. Further reductions in coupling Rabi frequency beyond $\Omega_C \sim 2\pi \times 0.9$ MHz are detrimental to the SNR at all RF E-fields.

In Figure 4, we investigate detection performance for pulses with different lengths under laser conditions that give the overall highest SNR, $\Omega_P = 2\pi \times 3.5$ MHz and $\Omega_C = 2\pi \times 2.1$ MHz. We evaluate the timing precision as the standard deviation of a Gaussian fit to a distribution of 300 pulse timings extracted in post-processing from the matched filter maxima. Timing precision and SNR for weak RF fields are limited by small pulse depth. At larger RF E-fields timing precision can be limited by the downsampling required to implement the matched filter on the FPGA. Limits from sampling rate are indicated as horizontal dashed lines in Figure 4b, which generally can be increased for shorter pulses.

If the EIT peaks are fully split then the SNR remains flat with varying RF E-field, but drops once overlap occurs at weaker RF E-fields. At a SNR ~15 dB we begin to see rare false alarms (<1%), which increase at lower SNR. Pulses shorter than 0.5 μs do not have time to reach sufficient depth because of the finite atomic response time (Figure 1d), which degrades the SNR and timing precision at all RF amplitudes. Despite this, we are able to detect individual short pulses with the matched filter, down to the RF generator's 50 ns width limit.

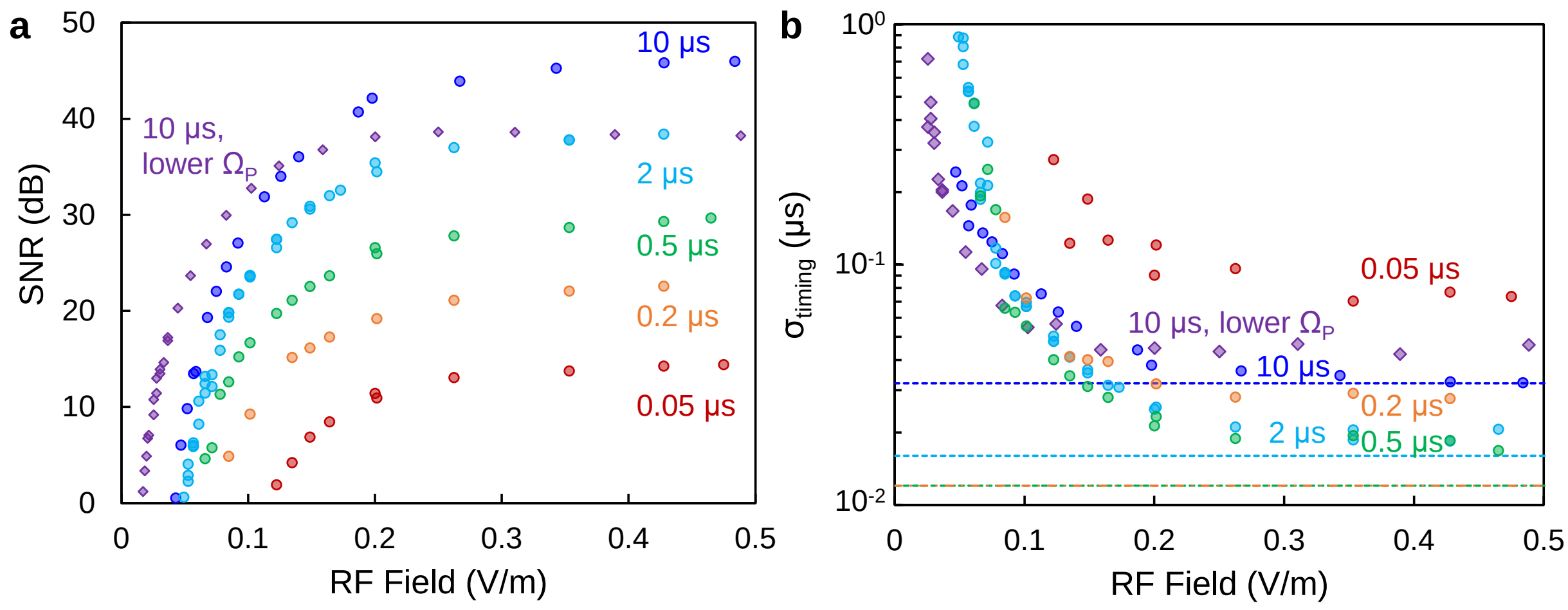

**FIG. 4 –** (a) Signal-to-noise (SNR) ratio of the matched filter shown for various RF pulse lengths (circles) and 10 μs pulses taken at a lower IR power ($\Omega_P = 2\pi \times 1.7$ MHz instead of $\Omega_P = 2\pi \times 3.5$ MHz). (b) Timing precision for the various pulse conditions in panel (a), taken as the standard deviation of a Gaussian fit to pulse timings measured from the matched filter peak. Dashed horizontal lines correspond to limits due to the finite sampling rate of the FPGA implementing the matched filter.

Using optimal laser conditions for detecting weak RF pulses (purple), $\Omega_P = 2\pi \times 1.7$ MHz and $\Omega_C = 2\pi \times 2.1$ MHz, our RF E-field sensitivity reaches ~170 μV cm$^{-1}$ when the SNR ~ 0 dB. For a 2 μs sensing time, this corresponds to a sensitivity of approximately 240 nV cm$^{-1}$ Hz$^{-1/2}$. When the SNR approaches 15 dB and occasional false alarms become apparent, corresponding to an RF E-field limit of ~330 μV cm$^{-1}$, a 2 μs sensing time yields an effective sensitivity of 470 nV cm$^{-1}$ Hz$^{-1/2}$. These sensitivities are obtained in real time on single pulses, without the need for an auxiliary reference RF E-field, averaging, or additional modulation.

## Use of Burst RF Pulse Sequences

We also investigate the application of the FPGA-implemented matched filter to short bursts of RF pulses in order to improve detection. We use a series of three 2 μs pulses spaced by 2 μs, so that the total pattern is 10 μs long but contains lower total energy than a single 10 μs pulse. The template used for the matched filter is shown in Figure 5a, and the output on a typical noisy experimental pulse is given in Figure 5b. The burst pattern results in a SNR lower than a single 10 μs pulse at large RF E-field amplitude due to the lower total energy, but maintains a similar SNR at weak RF E-fields. The burst pattern produces a narrower peak in the matched filter output, which improves timing precision at low to moderate RF E-fields and exceeds that of either a single 2 μs or 10 μs pulse. However, we find that due to the addition of sidelobes in the matched filter pattern the use of burst sequences comes with an added false alarm rate of ~5% resulting from the cases where the sidelobe exceeds the central peak due to noise. Improvements could be achieved by varying the amplitude of each pulse in the sequence based on compression techniques [25,26].

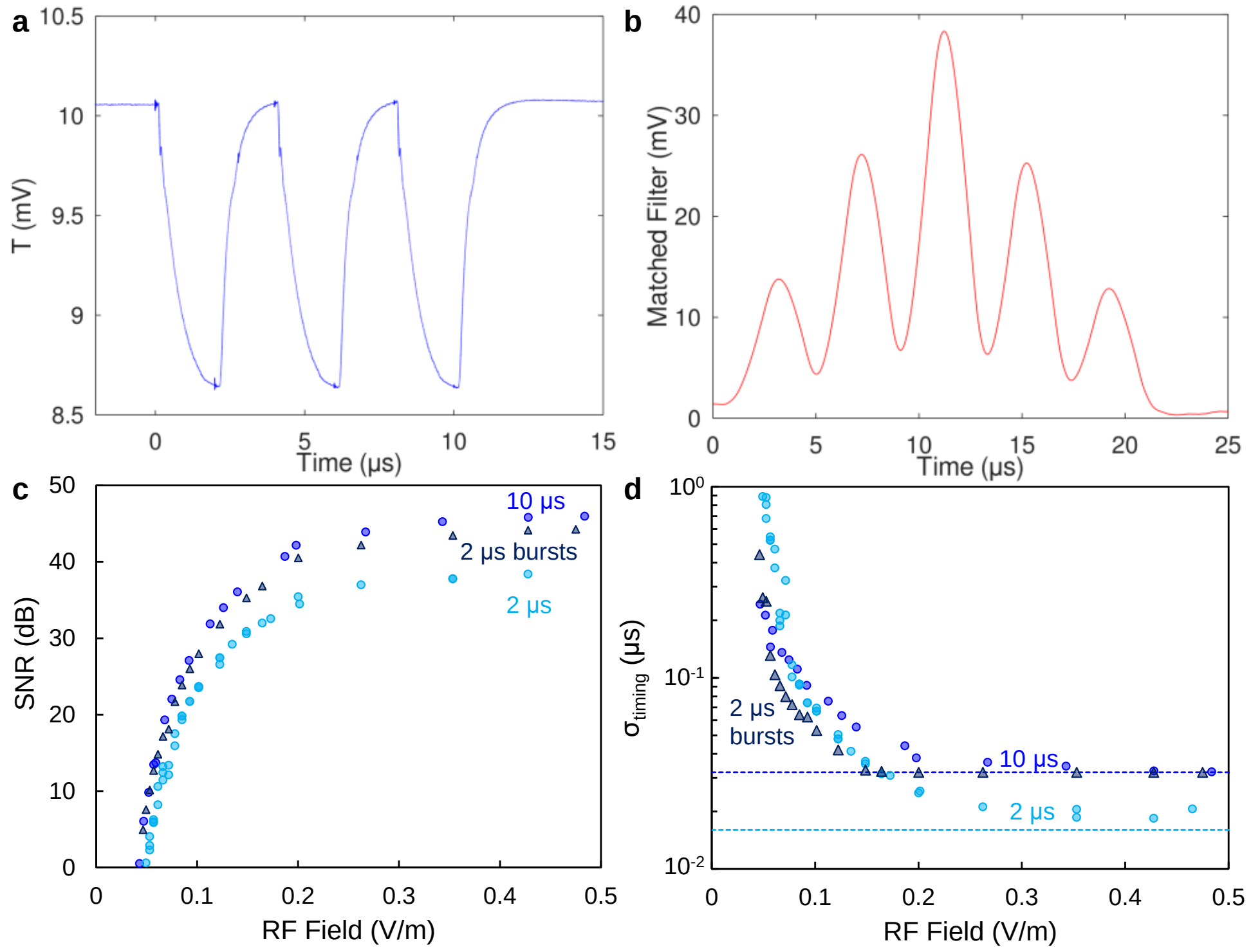

**FIG. 5 –** Use of a burst sequence to improve pulse detection. (a) An averaged atomic response to a series of three 2 µs RF pulses spaced by 2 µs, forming the template for the matched filter. (b) Application of the matched filter to a single noisy burst sequence produces an output with a narrower central peak that is less prone to inaccuracy at weak RF E-fields, but with additional sidelobes that lead to false alarms. (c) Bursts (dark blue triangles) show an SNR between that of a single 10 µs (dark blue circles) and single 2 µs pulse (light blue circles). (d) Burst sequences demonstrate improved timing accuracy over single pulses.

## Nonlinearity and Matched Filter Template Mismatch

Using the matched filter implies a linear time-invariant system, but the atomic system is nonlinear and time-varying. Given that the template is by necessity pre-assigned and unable to respond in real-time to changing conditions, the consequence is that there may be mismatch between the pulse shape and template, altering the extracted pulse arrival time and SNR.

In Figure 6a, we examine how the pulse shape changes nonlinearly with RF E-field amplitude. For a large range of RF E-fields, we find that the pulse shape remains largely the same, with only a subtle change to the slower time constant after the initial transient. However, once the amplitude regime is reached at low RF E-fields, below ~0.1 V $m^{-1}$, both edges of the pulse develop overshoots due to Rabi oscillations associated with the fast transient. Variations of evenness in the pulse between ~2.5 µs to 10 µs are due to amplitude fluctuations of the RF generator. The atoms are more sensitive to these variations at weak RF E-fields, i.e. in the sloped regions of Figure 3c.

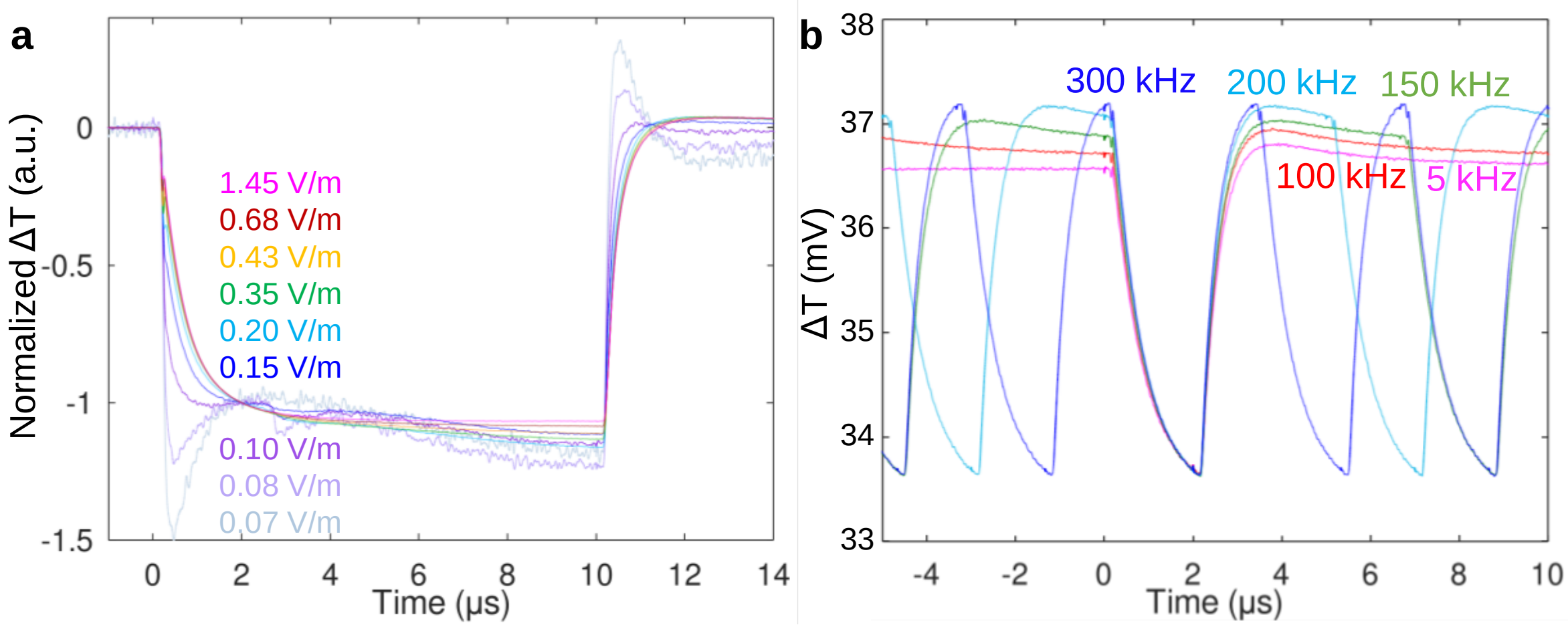

**FIG. 6 –** (a) Pulse shape at different RF E-fields, normalized to the same depth. At weaker fields, the atomic response is slightly more rapid, and as the RF E-field amplitude is further decreased overshoots begin to develop on each edge. $\Omega_P = 2\pi \times 3.5$ MHz, $\Omega_C = 2\pi \times 2.1$ MHz. (b) Effects of repetition rate on pulse shape for 2 μs long pulses. $\Omega_P = 2\pi \times 8.8$ MHz, $\Omega_C = 2\pi \times 2.1$ MHz, $\Omega_{RF} = 2\pi \times 120$ MHz.

The shape of the atomic response can also vary with the pulse repetition rate due to collisions, ionization, and electric field effects with substantially longer equilibration timescales than a single pulse. Figure 6b shows that the pulse depth and time scales vary slightly with pulse rate. In particular, when the pulse repetition rate is low and dark state atoms or ions have time to accumulate, then an overshoot is present on the tail of the pulse.

In Figure 7, we demonstrate what happens if the matched filter template (red inset) is not a perfect match to the underlying pulse shape being detected (blue inset). This can arise primarily from the nonlinearity and time variance of the atomic system, but may also occur due to changes in the vapour cell itself (*e.g.* electric fields, vapour pressure) or the lasers (*e.g.* power, frequency detuning). For this analysis, we perform matched filter computations in post-processing on averaged data rather than implemented for real-time analysis on the FPGA hardware to avoid shifts due to noise. We find that differences in the time scales comprising the pulse, or the presence of overshoots on the edges, result in slight shifts to the pulse time extracted at the maximum of the filter output that are generally <0.5 μs for a 10 μs long pulse. The direction of the shift can be later or earlier, depending on the nature of the pulse-template mismatch. The symmetric presence of overshoots on the leading and trailing edges of the pulse helps mitigate any shift due to differences in RF amplitude, but detection could be improved by using two filters run in parallel, one designed for large RF amplitudes and another for small RF amplitudes. On the other hand, temporal variation can be minimized by working in a weak field EIT limit, i.e. with Rabi frequencies chosen for low Rydberg state populations.

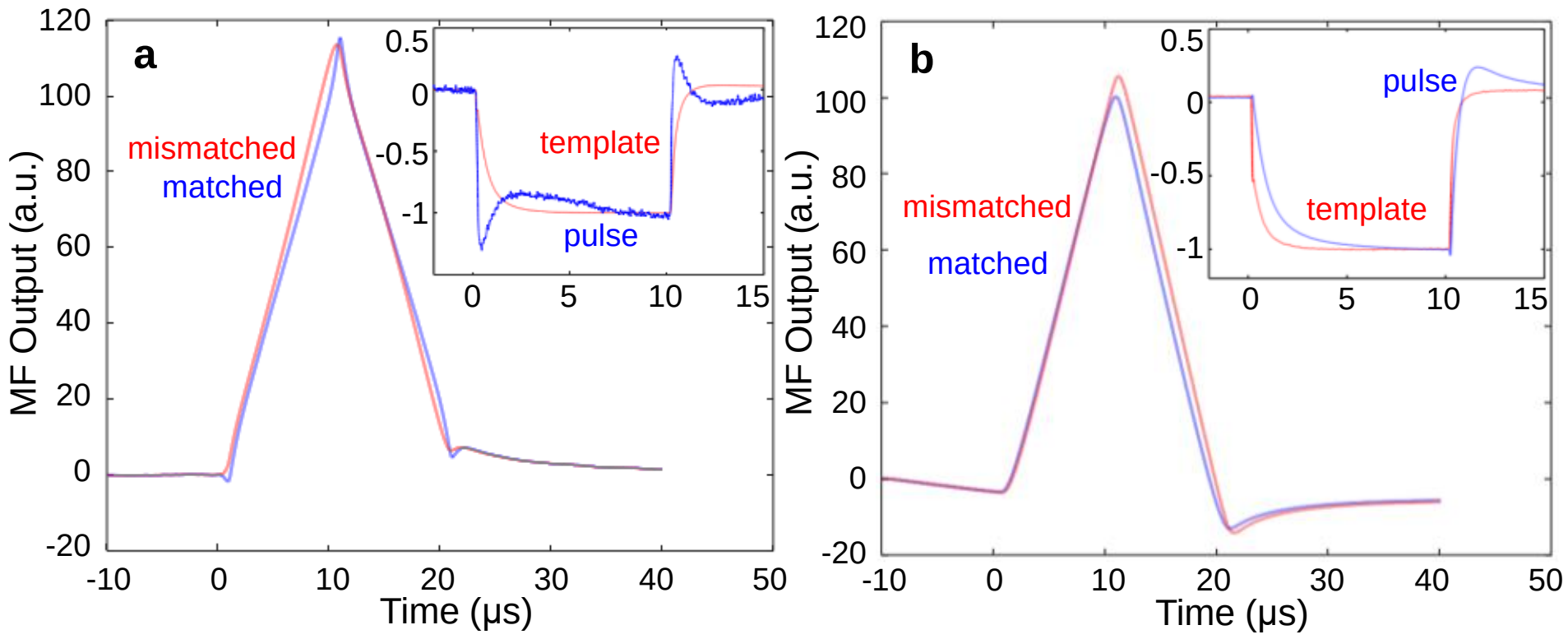

**FIG. 7 –** Using a template that does not fully match the atomic pulse shape produces slight shifts in the extracted pulse timing from the matched filter (red) compared to using an ideal match (blue). The insets show the shapes of the template (red) and target pulse (blue). (a) A pulse template for large RF E-field amplitudes is used on a weak RF E-field atomic response. (b) A scenario where the target pulse contains unanticipated strong collisional, ionization, and electric field effects on the tail.

## Radar Detection

Finally, we demonstrate the experimental detection of emulated radar signals from an aircraft flying past our vapour cell receiver with a path shown in Figure 8a. The signal is produced by a vector signal generator that simulates the power and timing of pulses that reach the receiver accounting for changing distance, which we send to the RF horn antenna near the vapour cell. The simulated aircraft travels at a typical commercial airplane speed of 200 m/s at a distance of around 1 km from the receiver. On board is a spinning antenna rotating at 30 pm, emitting 1 μs long pulses at a repetition rate of 10 kHz. The simulated peak power output is around 6 dBm, which is delivered to the RF horn antenna. For detection, we use laser conditions corresponding to the highest RF E-field sensitivity, with $\Omega_P = 2\pi \times 1.7$ MHz and $\Omega_C = 2\pi \times 2.1$ MHz.

Figure 8b shows the radar pattern seen through the atomic response of the vapour cell, processed with the FPGA-implemented matched filter. Clusters of pulses are visible when the emitter faces the receiver, whose amplitudes vary in a typical antenna pattern (Figure 8c). Due to the atomic nonlinearity and saturation at high RF amplitudes (see Figure 3c), the central lobe appears flat rather than rounded like the sidelobes. Furthermore, the first sidelobe has a slightly higher amplitude than the central lobe due to the enhanced absorption on the sides of the EIT peak at these laser conditions. As the aircraft approaches the receiver, the RF E-field at the vapour cell increases and more sidelobes can be distinguished above the noise floor. To detect a target from further away or distinguish weaker RF amplitudes, an increase in emitted pulse length, an increase in vapour cell depth, addition of an amplifier at the receiver (e.g. a dish), a reduction in noise sources, or averaging over several pulses is required.

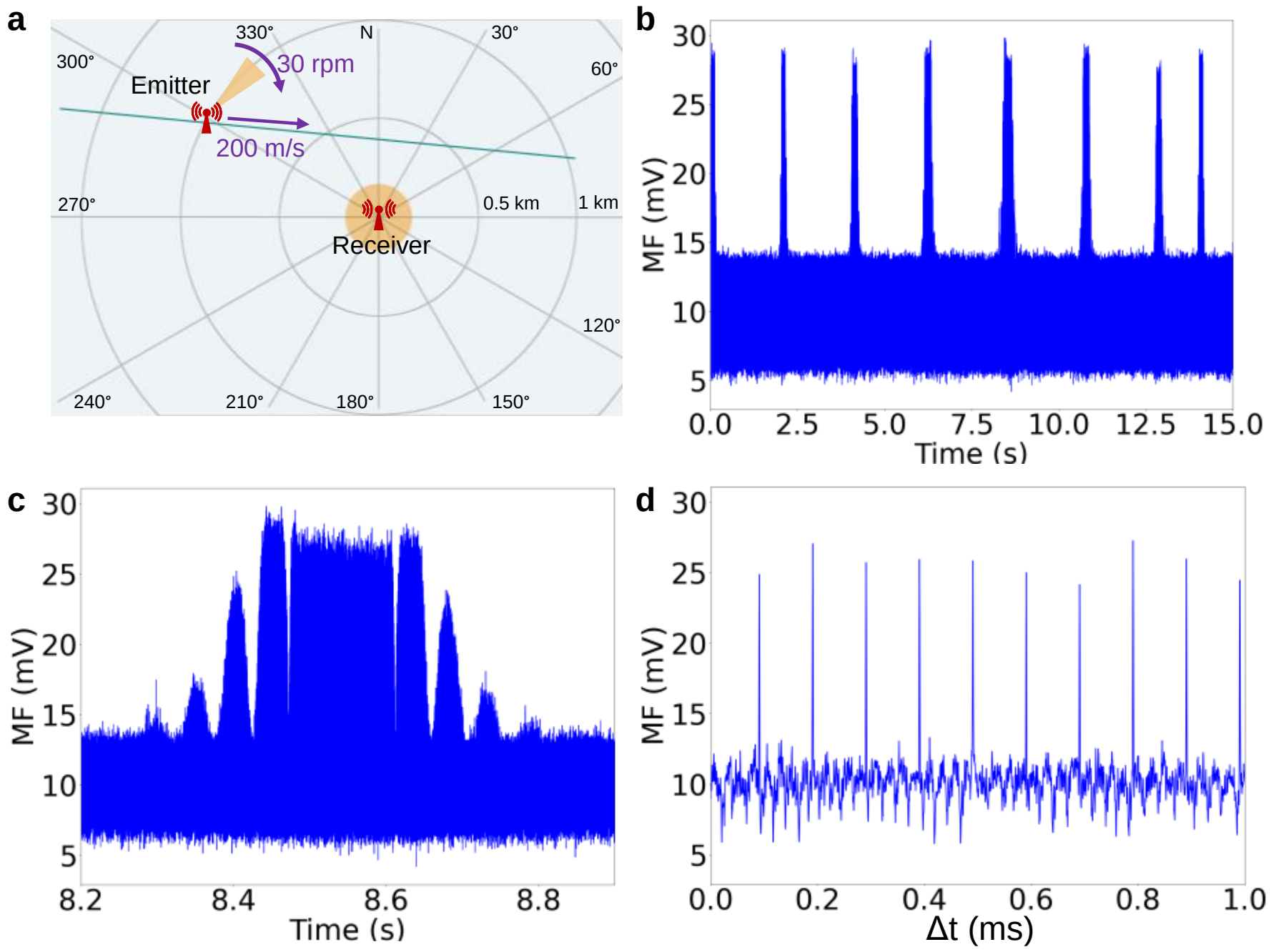

**FIG. 8 –** Radar detection. (a) Scenario describing a rotating emitter on an aircraft as it flies by a fixed vapour cell receiver, emitting 1 μs pulses at a repetition frequency of 10 kHz. (b) Detected signal after matched filtering, producing clusters of pulses whenever the emitter's antenna rotates to face the receiver. (c) Zooming into a pulse cluster shows the pattern formed from a single rotation of the antenna, consisting of a large central peak with multiple weaker sidelobes. (d) Zooming further into the central lobe at 8.5 s shows the matched filter output from individual 1 μs pulses. $\Omega_P = 2\pi \times 1.7$ MHz, $\Omega_C = 2\pi \times 2.1$ MHz.

## Conclusion

In summary, we have shown that a room temperature Cs vapour cell has a rapid response to pulse-modulated RF E-fields and can detect RF pulses down to sub-50 ns durations. The overall atomic response time consists of two timescales: a short ~100 ns transient followed by a longer microsecond decay that depends on the transit time of Rydberg atoms and collisional by-products out of the interaction region. Using a matched filter based on the atomic response, we find that the vapour cell can detect single shot RF pulses down to amplitudes of ~170 μV $cm^{-1}$, with a sensitivity of ~240 nV $cm^{-1}$ $Hz^{-1/2}$ and timing precision of ~30 ns without use of an auxiliary field or other signal processing. We find that weak laser conditions are optimal for the highest sensitivity to low RF amplitudes, while higher laser powers are preferred for distinguishing a broader amplitude range. We demonstrated the functionality of Rydberg atom-based sensors as a radar receiver detecting pulses emitted by an aircraft antenna.

# Supplemental Material: The Origins of Rydberg Atom Electrometer Transient Response and its Impact on Radio Frequency Pulse Sensing

Stephanie M. Bohaichuk, Donald Booth, Kent Nickerson, Harry Tai, James P. Shaffer

Quantum Valley Ideas Laboratories, 485 Wes Graham Way, Waterloo, ON N2L 6R2, Canada

### *Estimating the Dark State Generation Rate*

In our density matrix model, $\Gamma_{35}$ represents a dark state generation rate for atoms and/or ions, arising primarily through Rydberg-Rydberg collisions, which is important at optical conditions which generate high Rydberg state populations. We can obtain an estimated value for this rate for typical experimental values using [1]:

$$\Gamma_{35} \sim \eta N_g \rho_{33} \bar{v} \sigma \tag{S1}$$

Where $N_g \sim 3\times10^{16}\mathrm{m}^{-3}$ is the number density of Cs atoms in the vapour cell at room temperature, $\rho_{33}\sim0.1\%$ is an approximate population in the Rydberg state $|3\rangle$ for our typical experimental conditions, $\bar{v}$ =219 m/s is the mean velocity of atoms, and $\sigma = \pi R_{nn}^2$ is the cross-section for Rydberg-Rydberg collisions. Using pair-wise interaction potentials for two $55D_{5/2}$ state atoms [2], we estimate the internuclear distance at which a collision interaction occurs to be $R_{nn} \sim 7$ μm, corresponding to a shift in the energy levels equivalent to the laser spectral bandwidth. If we assume that one in every ten collisions produces an atom or ion in a dark state, $\eta = 0.1$, then we obtain an estimated collision rate on the order of $\Gamma_{35} \sim 2\pi\times100$ kHz, which is the same order of magnitude as the values obtained from fitting the experimental data. Note that the shifted atoms are effectively in a dark state as well due to blockade, but that Doppler shifts also play a role in determining which atoms participate in the EIT process.

We also expect there to be some ions generated in the vapour cell due to, for example, Penning ionization or blackbody radiation. At the typical Rydberg densities we work with, blackbody radiation induced ionization rates are expected to be several hundred Hz [3], while Penning ionization rates are lower [4,5]. While the generation rate of ions is expected to be very low, their long-range interactions can still have an impact on atoms in a Rydberg state. Here, we provide simple estimates for the plausibility of ionization influencing the pulse shape. Based on calculations for the $55D_{5/2}$ state [2], we expect that a >10 mV/cm electric field is needed to produce an appreciable Stark shift of >100 kHz. We estimate that a Rydberg atom must therefore be within $R_{ion}\sim38$ μm of the ion to experience the required electric field, given the $1/r^2$ dependence of an electric field around a charged particle. Similar to Equation S1, we can write an approximate rate $\Gamma_{ion}$ for the disturbance of a given Rydberg atom via interaction with an ion, *i.e.*, how often the Rydberg atom ends up within $R_{ion}$ of an ion:

$$\Gamma_{ion} \sim N_g \rho_{ion} \bar{v} \sigma, \tag{S2}$$

where $\rho_{ion}$ is the ion population and $\sigma = \pi R_{ion}^2$ is an estimated cross-section for Rydberg-ion interactions. To obtain an ion population in the long time between pulses, when the RF is off, we can solve the following rate equation in steady state:

$$\frac{dn_{ion}}{dt} = \rho_{33} N_g \mathrm{G}_{ion} - n_{ion}\Gamma_{ti} \sim 0, \tag{S3}$$

where $n_{ion} = \rho_{ion} N_g$ is the number density of ions, $\Gamma_{ti}$ is the transit time of ions through the interaction region, and $G_{ion}$ is the generation rate of ions from blackbody radiation and Penning ionization. We expect the effective interaction region width to be larger than the laser beams because of the large effective radius of an ion's influence, $R_{ion}$, which results in a slower transit time $\Gamma_{ti} \sim 2\pi\times175$ kHz. Solving Equation S3 and S2, we expect roughly 0.2% of Rydberg atoms to end up as ions, resulting in a disturbance rate with an order of magnitude around $\Gamma_{ion} \sim 2\pi\times150$ kHz which is similar to our Rydberg-Rydberg collision rate estimate. Again, we obtain a value that is the same order of magnitude as obtained from fits to the data.

In practice, we find a range of dark state population rates from $\Gamma_{35} \sim 0$ to $2\pi\times800$ kHz produce better fits of our model to the data, with larger rates corresponding to experimental conditions where the Rydberg population is larger. For larger Rydberg state populations, background electric fields in the cell, collisions, and ions may all play a role in the dynamics and are expected to increase the effective $\Gamma_{35}$ rate. Collisions and ionization are difficult to quantitatively predict in vapour cells, and in general atomic pulse shapes at higher optical Rabi frequencies will require fitting of the dark state generation rate to capture these complex effects, as shown in Figure S1. All in all, the estimates presented here for the collisional effects plausibly support the $\Gamma_{35}$ rates used in the model.

We note that in our study the Rydberg state $|3\rangle$ is primarily involved in dark state generation, not the upper Rydberg state $|4\rangle$. The atomic response is primarily shaped by the dark state population accumulated before the pulse turns on, when the RF is off and no upper state $|4\rangle$ population is present. Once the RF pulse is applied, population can be produced in $|4\rangle$ but does not meaningfully contribute to the dark state population because of the short time over which it is coupled to the optical system. Neglect of the $|4\rangle$ population may not be accurate for other pulse lengths and repetition rates.

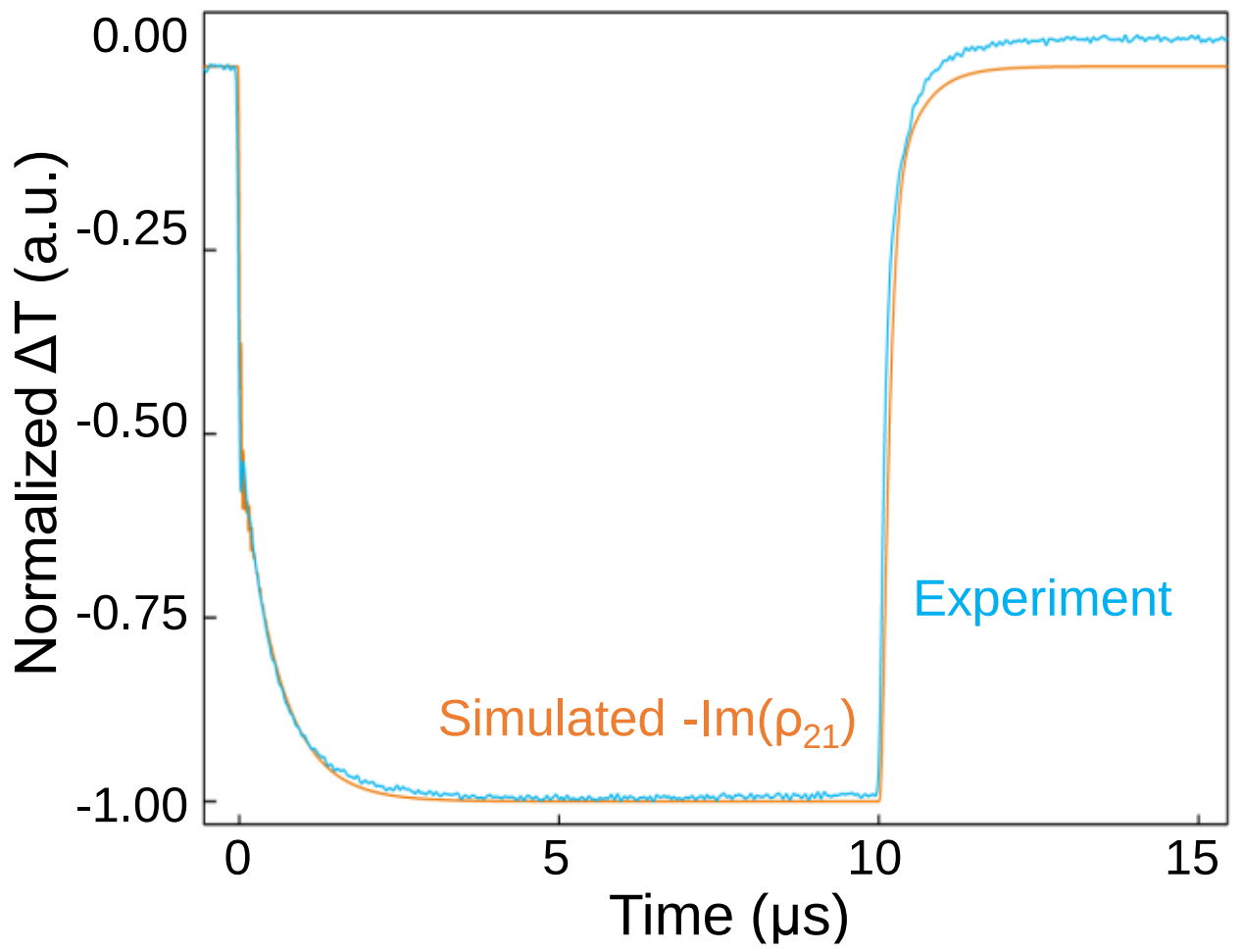

**FIG. S1** – Experimental atomic response to a 10 μs RF pulse (blue) compared to one simulated with a density matrix model (orange) at conditions with higher Rydberg populations. The presence of dark state atoms and/or ions is required in the model to better fit the slower pulse time scales at both the leading and trailing edges. An overshoot is present on the tail in experimental data, likely due to the presence of electric fields and ions, discussed below. $\Omega_P = 2\pi\times2.0$ MHz, $\Omega_C = 2\pi\times2.1$ MHz, and $\Omega_{RF} = 2\pi\times60$ MHz.

### *Pulse Time Scales*

We find that the second slower time constant making up the leading edge of the atomic pulse response is sensitive to the choice of laser beam radius, shown in Fig. S2a. Here we have varied the probe beam radius, which is smaller than the coupling laser beam, and maintained a constant Rabi frequency by increasing the probe laser power. The initial <100 ns transient present on the leading edge of pulses is unaffected by the laser beam size, to within the bandwidth resolution of the photodetector. However, a time constant can be extracted from an exponential fit to the second slower decay of the pulse, and this is found to scale monotonically with the infrared laser's radius as shown in Fig. S2b. As the transit time of the atoms through the beam is directly proportional to its radius, this demonstrates that this slower transient time scale of the pulse is dominated by the atomic transit time.

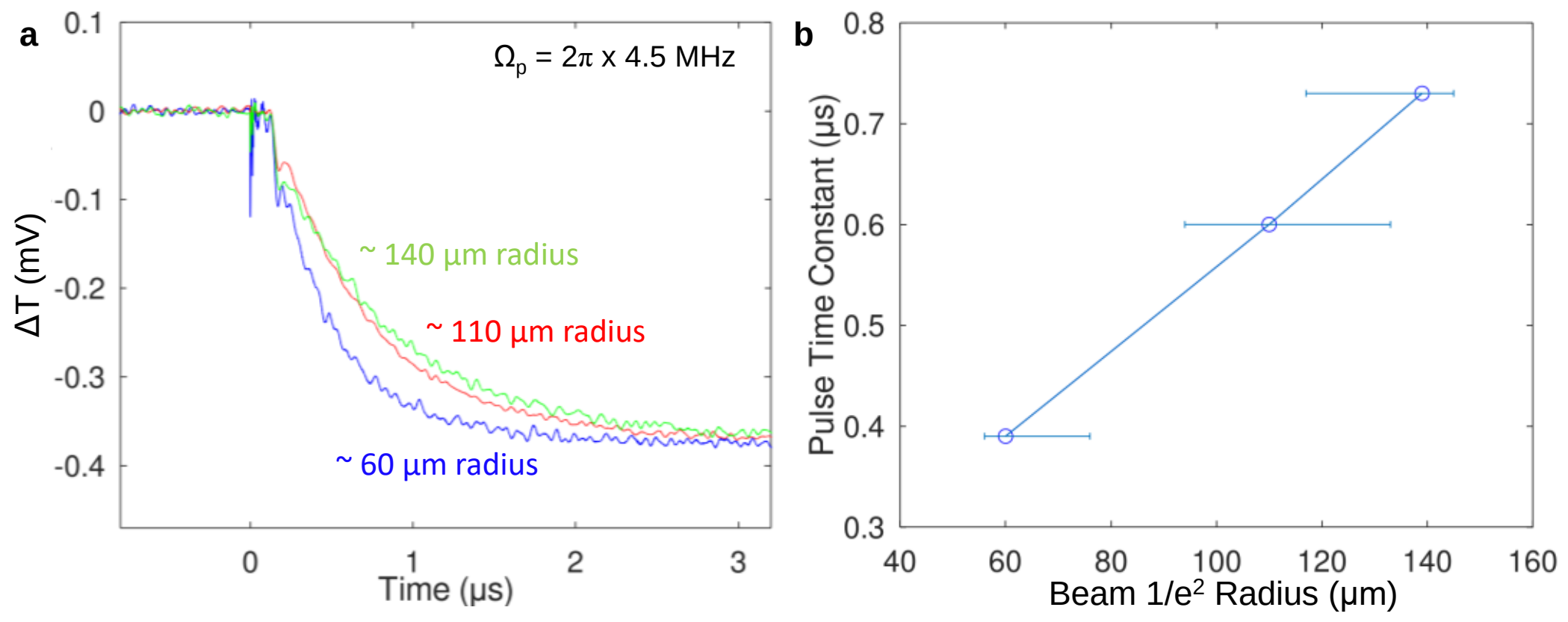

**FIG. S2** – Effects of laser beam size on pulse shape. (a) Increasing the probe beam radius slows the atomic response at the leading edge of the pulse. (b) The time constant extracted from an exponential fit to the leading edge of the pulse scales linearly with the probe beam radius.

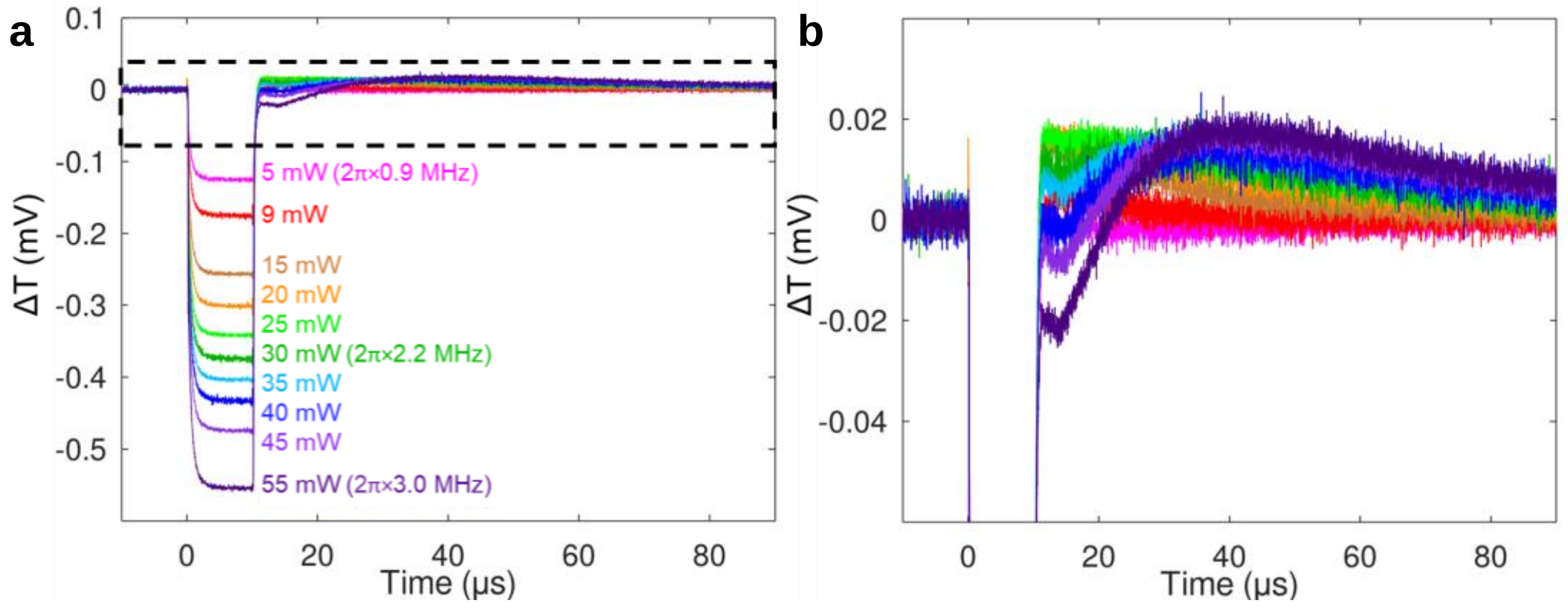

**FIG. S3 -** Electric field effects. (a) The trailing edge of the pulse depends on the coupling laser's Rabi frequency. A stronger overshoot is present at higher Rabi frequencies, whose effects can linger over many tens of microseconds. $\Omega_P = 2\pi \times 1.8$ MHz (b) Magnification of the dashed box in (a).

The shape of the trailing edge depends on the coupling laser power, shown in Fig. S3. An increase of the coupling laser power tends to increase the presence of overshoots and long transients on the tail due to the increase in Rydberg state population and therefore in the dark state atom and ion population. We also find that the shape and time scale of the trailing edge can depend on the laser position relative to the vapour cell walls, shown in Fig. S4. This suggests the presence of a non-zero background DC electric field in the cell, which is typically stronger closer to the walls due to surface charges. The depth of the pulse changes as the laser positions are shifted primarily due to changes in optical transmission through the glass faces of the vapour cell. All of these effects highlight that the slow trailing edge of the pulse can be sensitive to factors such as collisions, generation of ions, and background electric fields in the vapour cell.

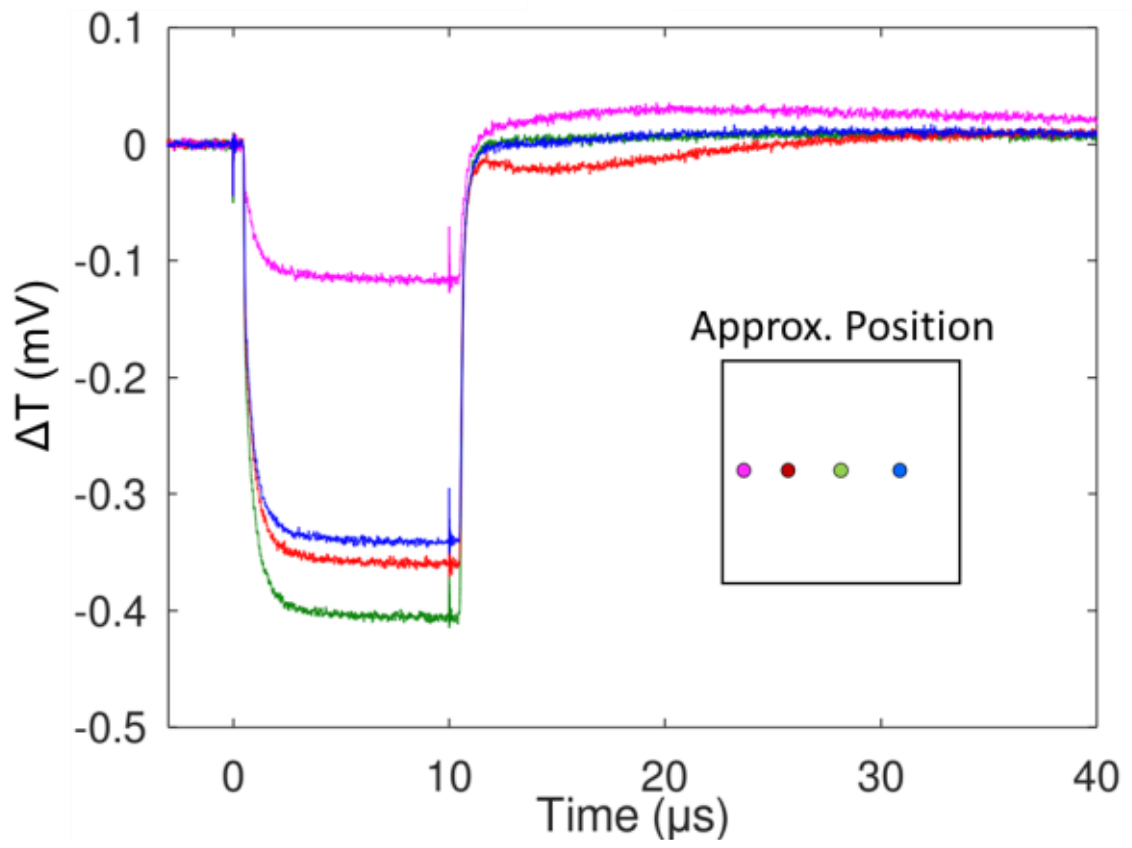

**FIG. S4 –** Electric field effects. While the shape of leading edge of the pulse does not change with the laser beam positions relative to the vapour cell walls, the shape of the trailing edge does. The inset shows an approximate location for the laser beams within the 1 cm x 1 cm face of the vapour cell, with the lasers going in and out of the page. $\Omega_P = 2\pi \times 1.8$ MHz, $\Omega_C = 2\pi \times 2.6$ MHz.

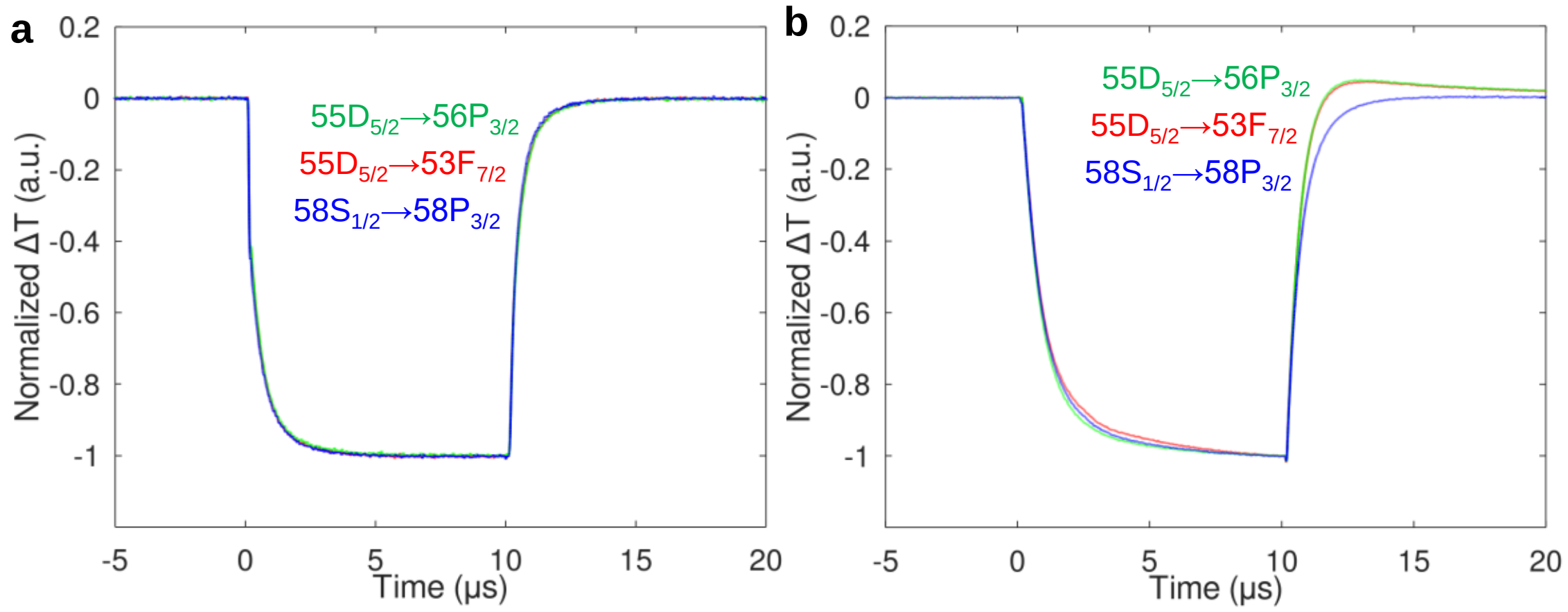

**FIG. S5** – Effects of atomic transition choice on experimental pulse shape. (a) At low optical Rabi frequencies ($\Omega_P = 2\pi \times 1.8$ MHz, $\Omega_C = 2\pi \times 0.6$ MHz) changes to the Rydberg transitions have no visible impact on the leading edge of the atomic response. (b) At higher probe laser power, $\Omega_P = 2\pi \times 12.2$ MHz, the choice of Rydberg transitions has minimal effect on the leading edge of the pulse, but does affect the trailing edge, likely due to the varying collisional rates and polarizability among states. $\Omega_{RF} = 2\pi \times 60$ MHz.

Finally, in Fig. S5 we investigate the role that the Rydberg state has on the pulse shape. In addition to the $55D_{5/2} \leftrightarrow 53F_{7/2}$ transition (K-band 19.40 GHz microwaves), we examine the $55D_{5/2} \leftrightarrow 56P_{3/2}$ transition (C-band 4.2 GHz microwaves) and the $58S_{1/2} \leftrightarrow 58P_{3/2}$ transition (K-band 18.9 GHz microwaves, with the coupling laser shifted to 509.26 nm) at large RF E-fields when the EIT peaks are fully split. For the C-band measurements, we change to a ~4 GHz RF horn antenna placed ~40 cm from the cell. However, we maintain fixed Rabi frequencies for all three transitions. Qualitatively, the pulse shape is identical for all the different choices of atomic transitions at low Rabi frequencies (*i.e.*, low laser powers) when Rydberg collision and ionization rates are low, which is optimal for weak RF pulse detection. We expect to be able to sense pulses with similar temporal accuracy using different atomic states, albeit with different degrees of sensitivity due to the different dipole moments of the states, $\mu_{34}$.

At higher laser powers, we observe different degrees of overshoot on the trailing edge of the pulses, though the leading edge remains the same. The $58S_{1/2} \leftrightarrow 58P_{3/2}$ transition has equal pre- and post-pulse transmission levels, appearing to avoid significant collisional and electric field effects, unlike the other transitions. To understand this, we calculate pair-wise potentials for the $58S_{1/2}$ and $55D_{5/2}$ Rydberg pair states shown in Figure S6. These calculations include both dipole and quadrupole interactions [2]. All of the interaction potentials for the $58S_{1/2}$ state are repulsive, and the steep avoided crossings shown are unlikely to result in transitions due to extremely low Landau-Zener curve crossing probabilities. The crossing of the $58S_{1/2}$ pair state with the nearby $55F_{7/2}$+$53G_{9/2}$ is not directly allowed by either the quadrupole or dipole interactions. In contrast, the $55D_{5/2}$ pair state potentials include attractive potential curves that can more easily lead to state changing collisions. Furthermore, we calculate the polarizability of the $55D_{5/2}$, $m_J$=5/2 and the $58S_{1/2}$, $m_J$=1/2 states to be -2800 MHz/(V/cm)$^2$ and 155 MHz/(V/cm)$^2$ respectively. Given the significantly lower polarizability of the $58S_{1/2}$ state, we expect a lower sensitivity to Stark shifts driven by ionization and electric fields in the vapour cell. Since the $58S_{1/2}$ is more immune to both collisional and ionization effects, we see far less overshoot on the recovery of the pulse, indicating that there is less dark state produced when the RF pulse is off. The data showing that the results for the $56P_{3/2}$ and $53F_{7/2}$ Rydberg states are effectively identical supports our earlier argument that the RF coupled upper Rydberg state plays little role in populating the dark state. Overall, the change in pulse shape with Rydberg population, the form of the pair potentials, and the data shown in Fig. S5 support our explanation that the tail of the pulse is predominantly influenced by the population of dark states due to collisions, ionization and electric fields.

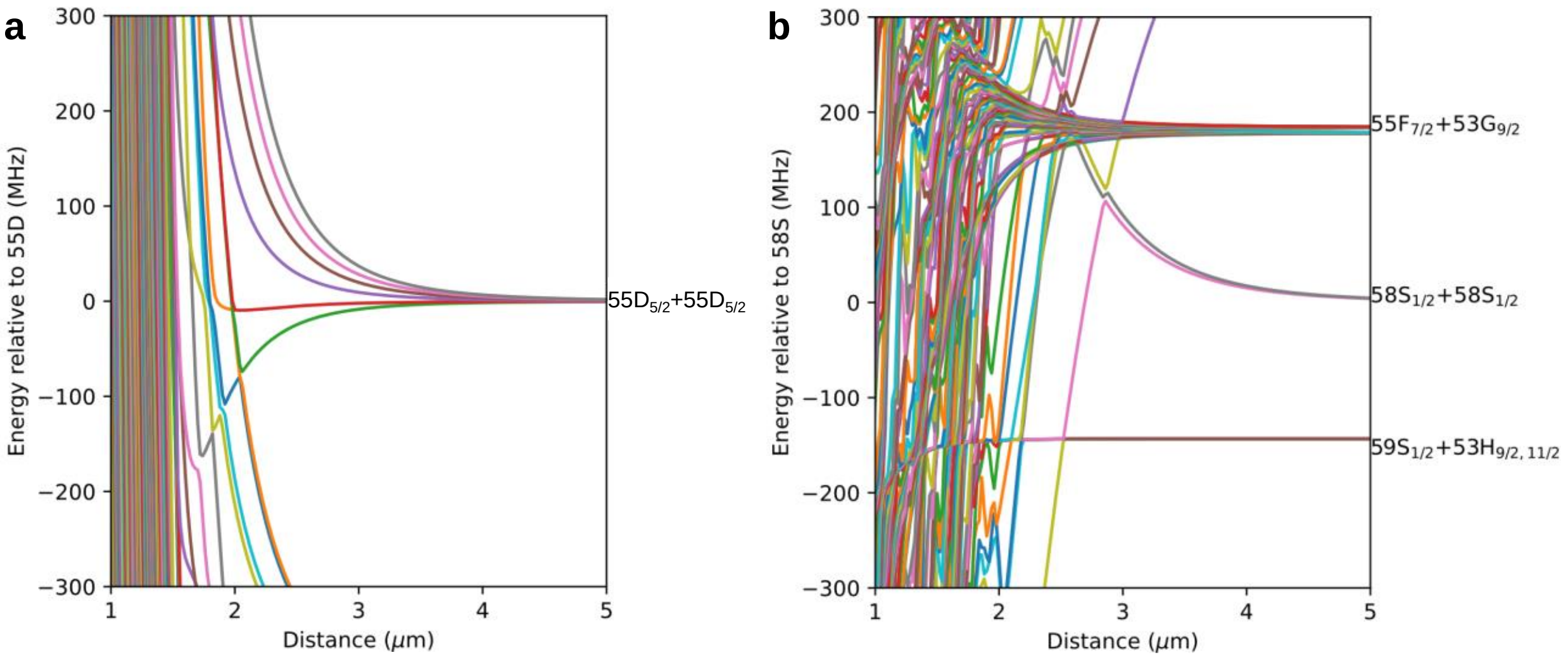

**FIG. S6 –** Pairwise interaction potentials calculated for (a) the $55D_{5/2}$ and nearby Rydberg states and (b) the $58S_{1/2}$ and nearby Rydberg states. The $55D_{1/2}$ pair states show some attractive interaction potentials that lead to collisions, unlike the $58S_{1/2}$ pair states which are all repulsive.